# Pilot Optimization and Channel Estimation for Multiuser Massive MIMO Systems

Tadilo Endeshaw Bogale and Long Bao Le
Institute National de la Recherche Scientifique (INRS)
Université de Québec,
Montréal, Canada
Email: $\{$tadilo.bogale, long.le$\}$@emt.inrs.ca

***Abstract*— This paper proposes novel pilot optimization and channel estimation algorithm for the downlink multiuser massive multiple input multiple output (MIMO) system with $K$ decentralized single antenna mobile stations (MSs), and time division duplex (TDD) channel estimation which is performed by utilizing $N$ pilot symbols. The proposed algorithm is explained as follows. First, we formulate the channel estimation problem as a weighted sum mean square error (WSMSE) minimization problem containing pilot symbols and introduced variables. Second, for fixed pilot symbols, the introduced variables are optimized using minimum mean square error (MMSE) and generalized Rayleigh quotient methods. Finally, for $N = 1$ and $N = K$ settings, the pilot symbols of all MSs are optimized using semi definite programming (SDP) convex optimization approach, and for the other settings of $N$ and $K$, the pilot symbols of all MSs are optimized by applying simple iterative algorithm. When $N = K$, it is shown that the latter iterative algorithm gives the optimal pilot symbols achieved by the SDP method. Simulation results confirm that the proposed algorithm achieves less WSMSE compared to that of the conventional semi-orthogonal pilot symbol and MMSE channel estimation algorithm which creates pilot contamination.**

## I. Introduction

Multiple input multiple output (MIMO) system is a promising approach for exploiting the spectral efficiency of wireless channels [1]–[3]. The MIMO system could be either a single user MIMO or multiuser MIMO.

The single user MIMO system requires expensive multi-antenna terminals at both the transmitter and receiver sides. Furthermore, one may not achieve the multiplexing gain of the single user MIMO channel when the propagation environment is poorly scattered [4]. In a multiuser MIMO system, the base station (BS) will serve several decentralized mobile stations (MSs). For such a system, employing multi-antenna at the BS and very cheap single antenna terminals at each MS appears to be the most economical design approach. And, in practice, as the MSs are sufficiently far apart from each other, the multiplexing gain of a MIMO channel can be ensured for multiuser scenario.

The multiuser MIMO system could be either in the uplink or downlink channel. For the downlink channel, it is recently shown in [5] that the full advantages of a MIMO system can be exploited using simple beamforming strategies such as maximum ratio transmission (MRT) or zero forcing (ZF), just by deploying very large number of antennas at the BS (i.e., Massive MIMO system) [4]. The basic idea is that when the number of BS antennas are large, the channel vector of any two MSs will be almost uncorrelated and the instantaneous channel gain of each MS can be well approximated (with high probability) from its second order statistics. However, to enjoy this benefit, the channel vector of each MS needs to be known (or estimated reliably) at the BS.

For the downlink multiuser MIMO system, channel estimation can be performed either in frequency division duplex (FDD) or time division duplex (TDD) approaches. In an FDD approach, first, the BS transmits a pilot sequence to all MSs. Then, each MS will estimate its own channel. Finally, the estimated channel is feedback to the BS via feedback channel. In a conventional channel estimation approach, the pilot sequences needs to be orthogonal [6]. To achieve this requirement, the BS requires $N$ symbol periods (which is significant if $N$ is very large). In the TDD system, first, each MSs transmits its own pilot symbols to the BS. Then, the BS will estimate the uplink channel and assign the conjugate of this estimated channel as its downlink version. Here also, the conventional orthogonal pilot design method requires $K$ symbol periods. In a massive MIMO system, as $K << M$, the TDD channel estimation method is the most reasonable approach [5]. However, when the number of MSs ($K$) is large, still allocating $K$ symbol periods for pilot transmission may not be practically feasible.

When $N < K$, one possible pilot design approach could be to reuse the pilots of the first $N$ MSs for the extra $K - N$ MSs. However, as will be seen later, such pilot reuse approach will create a so called pilot contamination. When there is a pilot contamination, the performance of the minimum mean square error (MMSE) channel estimator and downlink beamformer degrades drastically [5][1]. To alleviate the effect of pilot contamination, the Eigenvalue decomposition (EVD) based channel estimation approach is proposed in [7]. However, the complexity of this channel estimator is large when $M$ is very large which is the case in massive MIMO.

In the current paper, we propose novel pilot optimization and channel estimation algorithm for the downlink multiuser massive MIMO system with $K$ single antenna MSs and arbitrary $N$ pilot symbols (i.e., including the practically relevant $N < K$ pilot symbols), and TDD channel estimation method. The proposed algorithm is explained as follows. First, we formulate the channel estimation problem as a weighted sum mean square error (WSMSE) minimization problem containing pilot symbols and introduced variables. Second, for fixed pilot symbols, the introduced variables are optimized using minimum mean square error (MMSE) and generalized Rayleigh quotient methods. Finally, for $N = 1$ and $N = K$ settings, the pilot symbols of all MSs are optimized using semi definite programming (SDP) convex optimization approach, and for the other settings of $N$ and $K$, the pilot symbols of all MSs are optimized by applying simple iterative algorithm. When $N = K$, it is shown that the latter iterative algorithm gives the optimal pilot symbols achieved by the SDP method. Simulation results confirm that the proposed algorithm achieves less WSMSE compared to that of the conventional semi-orthogonal pilot symbol and MMSE channel estimation algorithm which creates pilot contamination.

[1]We would like to mention here that the pilot contamination in [5] is justified in the context of multi-cell MIMO system. However, the basic idea of pilot contamination is similar to that of the current paper.

This paper is organized as follows: Section II discusses the signal and massive MIMO channel model. In Section III, the conventional pilot assignment and channel estimation algorithm is presented. Our new pilot optimization and channel estimation algorithm is discussed in Section IV. In Section V, computer simulations are used to compare the performance of the proposed algorithm with the existing algorithm. Conclusions are drawn in Section VI.

*Notations:* In this paper, upper/lower-case boldface letters denote matrices/column vectors. $\mathbf{X}_{(i,j)}$, tr($\mathbf{X}$), det($\mathbf{X}$), $\mathbf{X}^H$ and E($\mathbf{X}$) denote the $(i,j)$th element, trace, determinant, conjugate transpose and expected value of $\mathbf{X}$, respectively. $\mathbf{I}_n$ is the identity matrix of size $n \times n$, $\mathcal{C}^{M\times M}$ and $\mathcal{R}^{M\times M}$ represent spaces of $M \times M$ matrices with complex and real entries, respectively. The acronym s.t and i.i.d denote "subject to" and "independent and identically distributed", respectively.

## II. Signal and Channel Model

We consider an uplink multiuser system where the BS equipped with $M$ antennas is serving $K$ decentralized single antenna MSs. It is assumed that $M >> K$ (massive MIMO system). During pilot transmission, the BS will receive the following signal at symbol time $t$

$$\mathbf{y}_t = \sum_{i=1}^{K} \mathbf{h}_i x_{it}^H + \mathbf{n}_t \tag{1}$$

where $\mathbf{h}_i \in \mathcal{C}^{M\times 1}$ is the channel between the $i$th MS and the BS, $x_{it}^H$ is the pilot symbol of the $i$th MS at period $t$ and $\mathbf{n}_t \in \mathcal{C}^{M\times 1}$ is the received noise vector at time slot $t$.

If we utilize $N$ symbol periods for pilot transmission, the received signal can be expressed in the following matrix form

$$\mathbf{Y} = \mathbf{H}\mathbf{X}^H + \mathbf{N} \tag{2}$$

where $\mathbf{x}_j = [x_{j1}, x_{j2}, \cdots, x_{jN}]^T$, $\mathbf{X} = [\mathbf{x}_1, \mathbf{x}_2, \cdots, \mathbf{x}_K]$, $\mathbf{H} = [\mathbf{h}_1, \mathbf{h}_2, \cdots, \mathbf{h}_K]$ and $\mathbf{N} = [\mathbf{n}_1, \cdots, \mathbf{n}_N]$.

The channel is modeled as $\mathbf{h}_k = [h_{k1}, h_{k2}, \cdots, h_{kM}]^T$ with $h_{km} = \sqrt{g}_{km}\tilde{h}_{km}, \forall m$, where $\sqrt{g}_{km}$ is the distant dependent path loss and $\tilde{h}_{km}$ is the fast fading part. It is assumed that $\tilde{h}_{km}$ is i.i.d random variable with $\tilde{h}_{km} \sim \mathcal{CN}(0,1), \forall k, m$, $\mathrm{E}\{\mathbf{N}_{(i,j)}\mathbf{H}_{(k,m)}\} = 0, \forall i,j,k,m$ and each element of $\mathbf{N}$ is also i.i.d zero mean circularly symmetric complex Gaussian (ZMCSCG) random variable all with unit variance. Furthermore, as the path loss depends on the distance between each MS and the BS, it is assumed that $\sqrt{g}_{km} = \sqrt{g}_k, \forall m$, and $\sqrt{g}_k, \forall k$ are known a priory.

## III. Conventional pilot assignment and MMSE channel estimation methods

In this section, we discuss the conventional pilot assignment and MMSE channel estimation algorithm of [6].

When $N = K$, the overall pilot symbol of all MSs ($\mathbf{X}$) is designed in such a way that $\mathbf{X}^H\mathbf{X} = \mathbf{I}_K$. The BS, then, decouples the channels of each MS by multiplying the overall received signal $\mathbf{Y}$ by $\mathbf{X}$, i.e.,

$$\begin{aligned}\mathbf{Z} &\triangleq \mathbf{Y}\mathbf{X} = \mathbf{H} + \mathbf{N}\mathbf{X} \\ &\Rightarrow \mathbf{z}_k = \mathbf{h}_k + \tilde{\mathbf{n}}_k, \ k = 1, \cdots, K\end{aligned} \tag{3}$$

where $\tilde{\mathbf{n}}_k = \mathbf{N}\mathbf{x}_k$ and $\mathbf{x}_k$ is the $k$th column of $\mathbf{X}$.

Now by introducing an MMSE receiver $\mathbf{W}_k^H$ for the $k$th MS, the estimated channel of the $k$th MS ($\widehat{\mathbf{h}}_k$) is expressed as

$$\widehat{\mathbf{h}}_k = \mathbf{W}_k^H \mathbf{z}_k, \forall k. \tag{4}$$

And $\mathbf{W}_k$ is designed such that the mean square error (MSE) between $\widehat{\mathbf{h}}_k$ and $\mathbf{h}_k$ is minimized. The MSE of the $k$th MS is given by

$$\begin{aligned}\boldsymbol{\xi}_k =& \mathrm{E}\{|\widehat{\mathbf{h}}_k - \mathbf{h}_k|^2\} \\ =& \mathrm{E}\{|\mathbf{W}_k^H(\mathbf{h}_k + \tilde{\mathbf{n}}_k) - \mathbf{h}_k|^2\} \\ =& \mathrm{E}\{|(\mathbf{W}_k^H - \mathbf{I}_M)\mathbf{h}_k + \tilde{\mathbf{n}}_k\} \\ =& g_k(\mathbf{W}_k^H - \mathbf{I}_M)(\mathbf{W}_k^H - \mathbf{I}_M)^H + \sigma^2\mathbf{I}_M.\end{aligned} \tag{5}$$

The optimal $\mathbf{W}_k$ is given as

$$\frac{\partial \boldsymbol{\xi}_k}{\partial \mathbf{W}_k^H} = 0 \Rightarrow \mathbf{W}_k^\star = \frac{g_k}{g_k + \sigma^2}\mathbf{I}.$$

When $K = 2N$ (i.e., $N < K$ scenario), the overall pilot symbols $\mathbf{X}$ can be designed as $\mathbf{X} = [\mathbf{U}, \mathbf{U}]$, where $\mathbf{U} \in \mathcal{C}^{N\times N}$ with $\mathbf{U}^H\mathbf{U} = \mathbf{I}_N$. With this settings[2], like in the above case, the BS will right multiply the received signal $\mathbf{Y}$ by $\mathbf{X}$ to get

$$\begin{aligned}\mathbf{Z} &\triangleq \mathbf{Y}\mathbf{X} = \mathbf{H} + \mathbf{N}\mathbf{X} \\ &\Rightarrow \mathbf{z}_k = \mathbf{h}_k + \mathbf{h}_{k+N} + \tilde{\mathbf{n}}_k, \ k = 1, \cdots, N \\ &\qquad = \mathbf{h}_k + \mathbf{h}_{K-k} + \tilde{\mathbf{n}}_k, \ k = N+1, \cdots, K.\end{aligned} \tag{6}$$

As we can see from this expression, the channel vector of the $k$th MS is coupled with the channel vector of the $(N+k)$ or $(K-k)$th MS. According to [5], this phenomena is called pilot contamination. When there is pilot contamination, the performance of the conventional MMSE channel estimation degrades drastically.

[2]Note that one can apply this pilot symbol design approach for any other settings of $K$ and $N$.

## IV. Proposed pilot optimization and channel estimation algorithm

In this section, we discuss our pilot optimization and channel estimation algorithm. To this end, we introduce the variables $\mathbf{u}_k$ and $\mathbf{W}_k$, and express the estimated channel of the $k$th MS as

$$\widehat{\mathbf{h}}_k = \mathbf{W}_k^H \mathbf{Y} \mathbf{u}_k, \ \forall k \tag{7}$$

where $\mathbf{W}_k \in \mathcal{C}^{M\times M}$ and $\mathbf{u}_k \in \mathcal{C}^{N\times 1}$ are the variables that will be determined in the sequel. Under these introduced variables, the MSE between $\widehat{\mathbf{h}}_k$ and $\mathbf{h}_k$ is given by

$$\begin{aligned}\boldsymbol{\xi}_k =& \mathrm{E}\{|\widehat{\mathbf{h}}_k - \mathbf{h}_k|^2\}\\ =& \mathrm{E}\{|\mathbf{W}_k^H(\sum_{i=1}^K \mathbf{h}_i\mathbf{x}_i^H + \mathbf{N})\mathbf{u}_k - \mathbf{h}_k|^2\}\\ =& \mathbf{u}_k^H(\sum_{i=1}^K g_i\mathbf{x}_i\mathbf{x}_i^H + \sigma^2\mathbf{I}_N)\mathbf{u}_k(\mathbf{W}_k^H\mathbf{W}_k) + g_k\mathbf{I}_M\\ &- (g_k\mathbf{x}_k^H\mathbf{u}_k)\mathbf{W}_k^H - (g_k\mathbf{u}_k^H\mathbf{x}_k)\mathbf{W}_k.\end{aligned}$$

It follows

$$\begin{aligned}\xi_k =& \mathrm{tr}\{\boldsymbol{\xi}_k\}\\ =& \mathbf{u}_k^H(\sum_{i=1}^K g_i\mathbf{x}_i\mathbf{x}_i^H + \sigma^2\mathbf{I}_N)\mathbf{u}_k\mathrm{tr}\{(\mathbf{W}_k^H\mathbf{W}_k)\} + g_k\mathbf{I}_M\\ &- (g_k\mathbf{x}_k^H\mathbf{u}_k)\mathrm{tr}\{\mathbf{W}_k^H\} - (g_k\mathbf{u}_k^H\mathbf{x}_k)\mathrm{tr}\{\mathbf{W}_k\}.\end{aligned}$$

As we can see from this expression, the MSE of each MS depends on its path loss. And in practice, we would like the channels of all MSs to be estimated reliably almost with the same accuracy. Due to this reason, we consider a WSMSE problem as our objective function, where the MSE weight of each MS is set to the inverse of its path loss. This optimization problem is mathematically formulated as

$$\begin{aligned}&\min_{\mathbf{x}_k,\mathbf{u}_k,\mathbf{W}_k} \sum_{k=1}^K \frac{1}{g_k}\xi_k\\ &\text{s.t } \mathbf{x}_k^H\mathbf{x}_k \leq P_k\end{aligned} \tag{8}$$

where $P_k$ is the maximum transmission power at each MS.

For fixed $\mathbf{u}_k$ and $\mathbf{x}_k, \forall k$, we can optimize $\mathbf{W}_k$ by the MMSE method as

$$\mathbf{W}_k = \frac{g_k\mathbf{x}_k^H\mathbf{u}_k}{\sum_{i=1}^K g_i\mathbf{x}_i^H\mathbf{u}_k\mathbf{u}_k^H\mathbf{x}_i + \sigma^2\mathbf{u}_k^H\mathbf{u}_k}\mathbf{I}_M. \tag{9}$$

Substituting this $\mathbf{W}_k$ back to $\xi_k$, we get the minimum MSE $\tilde{\xi}_k$ as

$$\tilde{\xi}_k = M\bigg(g_k - \frac{\mathbf{u}_k^H(g_k^2\mathbf{x}_k\mathbf{x}_k^H)\mathbf{u}_k}{\mathbf{u}_k^H(\sum_{i=1}^K g_i\mathbf{x}_i\mathbf{x}_i^H + \sigma^2\mathbf{I}_N)\mathbf{u}_k}\bigg). \tag{10}$$

From this expression, one can notice that $\mathbf{u}_k$ can be optimized independently by solving the following problem

$$\min_{\mathbf{u}_k} \tilde{\xi}_k = \max_{\mathbf{u}_k} \frac{\mathbf{u}_k^H(g_k^2\mathbf{x}_k\mathbf{x}_k^H)\mathbf{u}_k}{\mathbf{u}_k^H(\sum_{i=1}^K g_i\mathbf{x}_i\mathbf{x}_i^H + \sigma^2\mathbf{I}_N)\mathbf{u}_k}, \ \forall k. \tag{11}$$

Now by defining $\tilde{\mathbf{u}}_k \triangleq \mathbf{A}^{1/2}\mathbf{u}_k$ with $\mathbf{A} = \sum_{i=1}^K g_i\mathbf{x}_i\mathbf{x}_i^H + \sigma^2\mathbf{I}$, we can reformulate the above problem as

$$\max_{\tilde{\mathbf{u}}_k} \frac{\tilde{\mathbf{u}}_k^H\mathbf{A}^{-1/2}(g_k^2\mathbf{x}_k\mathbf{x}_k^H)\mathbf{A}^{-1/2}\tilde{\mathbf{u}}_k}{\tilde{\mathbf{u}}_k^H\tilde{\mathbf{u}}_k}.$$

This problem is the well known Rayleigh quotient optimization problem and its optimal solution is given by [8], [9]

$$\tilde{\mathbf{u}}_k^\star = \mathbf{A}^{-1/2}g_k\mathbf{x}_k \Rightarrow \mathbf{u}_k^\star = \mathbf{A}^{-1/2}\tilde{\mathbf{u}}_k^\star = \mathbf{A}^{-1}g_k\mathbf{x}_k. \tag{12}$$

Plugging this $\mathbf{u}_k^\star$ in (10) yields

$$\tilde{\tilde{\xi}}_k = Mg_k - Mg_k^2\mathbf{x}_k^H\mathbf{A}^{-1}\mathbf{x}_k. \tag{13}$$

By defining $\mathbf{V}_k \triangleq \mathbf{x}_k\mathbf{x}_k^H$, $\tilde{\mathbf{H}}_k \triangleq \sqrt{g}_k\mathbf{I}_N$, $\mathbf{V} = \mathrm{blkdiag}(\mathbf{V}_1, \cdots, \mathbf{V}_K)$ and $\tilde{\mathbf{H}} = [\tilde{\mathbf{H}}_1, \cdots \tilde{\mathbf{H}}_K]$, we can express $\sum_{k=1}^K \frac{1}{g_k}\tilde{\tilde{\xi}}_k$ as

$$\begin{aligned}\tilde{\tilde{\xi}}_w =& \sum_{k=1}^K \frac{1}{g_k}\tilde{\tilde{\xi}}_k\\ =& \sum_{k=1}^K M(1 - g_k\mathbf{x}_k^H\mathbf{A}^{-1}\mathbf{x}_k)\\ =& \sum_{k=1}^K M - M\mathrm{tr}\{(\tilde{\mathbf{H}}\mathbf{V}\tilde{\mathbf{H}}^H + \sigma^2\mathbf{I})^{-1}\tilde{\mathbf{H}}\mathbf{V}\tilde{\mathbf{H}}^H\}.\end{aligned}$$

If we apply matrix inversion lemma [10], we can rewrite $(\tilde{\mathbf{H}}\mathbf{V}\tilde{\mathbf{H}}^H + \sigma^2\mathbf{I})^{-1}\tilde{\mathbf{H}}\mathbf{V}\tilde{\mathbf{H}}^H$ as

$$(\tilde{\mathbf{H}}\mathbf{V}\tilde{\mathbf{H}}^H + \sigma^2\mathbf{I})^{-1}\tilde{\mathbf{H}}\mathbf{V}\tilde{\mathbf{H}}^H = \mathbf{I} - \sigma^2(\tilde{\mathbf{H}}\mathbf{V}\tilde{\mathbf{H}}^H + \sigma^2\mathbf{I})^{-1}.$$

Therefore, problem (8) can be equivalently formulated as

$$\begin{aligned}&\min_{\mathbf{V}_k} \ \mathrm{tr}\{(\sum_{k=1}^K g_k\mathbf{V}_k + \sigma^2\mathbf{I}_N)^{-1}\}\\ &\text{s.t } \mathrm{tr}\{\mathbf{V}_k\} \leq P_k,\\ &\quad \mathbf{V}_k \succcurlyeq \mathbf{0}, \ \mathrm{Rank}(\mathbf{V}_k) = 1, \forall k\end{aligned} \tag{14}$$

where $(.) \succcurlyeq \mathbf{0}$ denotes a PSD constraint. When $N = 1$ the rank constraint is always satisfied. And this solution is global optimal solution. Furthermore, when $N = K$, the optimal solution of the above problem can be obtained in closed form by applying the well known Majorization theory of [11] and is given by the conventional orthogonal pilot.

Now how can we solve this problem when $1 < N < K$ (i.e., the most practically relevant scenario). If we relax the rank constraint of this problem, we will get a standard convex SDP problem which can be solved efficiently with interior point methods with polynomial time complexity [8]. However, the drawback of this convex reformulation is that the optimal $\mathbf{V}_k$ is always a scaled diagonal matrix which is full rank. And we are not aware of any method to get the desired rank 1 solution from the solution of the above problem. Therefore, the critical issue will be how to get a rank one solution for the aforementioned settings of $N$ and $K$. In the following, we provide simple iterative algorithm that will give rank one solution for the above problem.

By employing matrix inversion lemma, we can re-express $(\sum_{k=1}^{K} g_k \mathbf{V}_k + \sigma^2 \mathbf{I}_N)^{-1}$ as [10]

$$
\begin{aligned}
(\sum_{i=1}^{K} g_i \mathbf{V}_i + \sigma^2 \mathbf{I}_N)^{-1} =& (\mathbf{Q}_k + g_k \mathbf{x}_k \mathbf{x}_k^H)^{-1} \\
=& \mathbf{Q}_k^{-1} - \frac{g_k \mathbf{Q}_k^{-1} \mathbf{x}_k \mathbf{x}_k^H \mathbf{Q}_k^{-1}}{1 + g_k \mathbf{x}_k^H \mathbf{Q}_k^{-1} \mathbf{x}_k} \\
\Rightarrow \mathrm{tr}\{(\sum_{i=1}^{K} g_i \mathbf{V}_i + \sigma^2 \mathbf{I}_N)^{-1}\} =& \mathrm{tr}\{\mathbf{Q}_k^{-1}\} - \frac{g_k \mathbf{x}_k^H \mathbf{Q}_k^{-2} \mathbf{x}_k}{1 + g_k \mathbf{x}_k^H \mathbf{Q}_k^{-1} \mathbf{x}_k}
\end{aligned}
$$

where $\mathbf{Q}_k = \sum_{i=1, i\neq k}^{K} g_i \mathbf{x}_i \mathbf{x}_i^H + \sigma^2 \mathbf{I}_N$.

Keeping $\mathbf{Q}_k$ constant, $\mathbf{x}_k$ can be optimized by solving the following problem

$$
\max_{\mathbf{x}_k} \frac{g_k \mathbf{x}_k^H \mathbf{Q}_k^{-2} \mathbf{x}_k}{1 + g_k \mathbf{x}_k^H \mathbf{Q}_k^{-1} \mathbf{x}_k}, \text{ s.t } \mathbf{x}_k^H \mathbf{x}_k \leq P_k. \quad (15)
$$

Since the objective function of this problem increases as $\mathbf{x}_k^H \mathbf{x}_k$ increases, at optimality $\mathbf{x}_k^H \mathbf{x}_k = P_k$ is satisfied. This problem can thus be reformulated as

$$
\max_{\mathbf{x}_k} \frac{g_k \mathbf{x}_k^H \mathbf{Q}_k^{-2} \mathbf{x}_k}{\mathbf{x}_k^H (\frac{1}{P_k} \mathbf{I}_N + g_k \mathbf{Q}_k^{-1}) \mathbf{x}_k}. \quad (16)
$$

The optimal solution of this problem can be obtained like in problem (11) and is given as $\mathbf{x}_k = \gamma_k \mathbf{F}_k^{-1/2} \tilde{\mathbf{x}}$, where $\tilde{\mathbf{x}}$ is the eigenvector corresponding to the maximum eigenvalue of the matrix $g_k \mathbf{F}_k^{-1/2} \mathbf{Q}_k^{-2} \mathbf{F}_k^{-1/2}$ with $\mathbf{F}_k = g_k \mathbf{Q}_k^{-1} + \frac{1}{P_k} \mathbf{I}_N$ and $\gamma_k$ is selected such that $\mathbf{x}_k^H \mathbf{x}_k = P_k$ is ensured.

The proposed pilot symbol design and channel estimation algorithm is summarized in **Algorithm I**.

**Algorithm I**

**Input parameters**: Set $N$, $K$, $\sigma^2$ and $g_k, \forall k$.

- ◇ Initialize $n = 0$ and $\mathbf{x}_k^n, \forall k$.

**Repeat**

- ◇ Using $\mathbf{x}_k^n, \forall k$, solve (16) for $k = 1, \cdots, K$ sequentially and update $\mathbf{x}_k^n, \forall k$ by the new $\mathbf{x}_k, \forall k$.
- ◇ Increase n by 1 (i.e., n=n+1).

**Until convergence**

- ◇ Compute $\mathbf{u}_k$ and $\mathbf{W}_k$ with (12) and (9), respectively.
- ◇ Using these $\mathbf{u}_k$ and $\mathbf{W}_k$, compute the estimated channel $\widehat{\mathbf{h}}_k$ with (7).

**Convergence analysis:** At each iteration, as the objective function of (16) (i.e., equivalent to (14)) is non-decreasing, this iterative algorithm is always convergent.

**Computational complexity:** As will be clear later in the simulation section, this algorithm converges in $n < 2$ iterations. Thus, the main computational cost of this algorithm arises from solving the Rayleigh quotient problem (16) which requires the computation of the eigenvalue decomposition of an $N \times N$ matrix. And in practice as $N << K << M$, the complexity of this algorithm is almost the same as the conventional pilot symbol and channel estimation algorithm discussed in Section III.

When $K = N$, it can be easily seen that this iterative algorithm will give the optimal solution of the SDP problem (14) when the initial $\mathbf{x}_k, \forall k$ satisfies $\mathbf{X}^H \mathbf{X} = \mathbf{I}_K$.

We would like to point out that the above pilot optimization and channel estimation algorithm can be applied for any $M$, $N$ and $K$. However, as discussed in [5] the effect of pilot contamination (i.e., pilot reuse) is worse for massive MIMO systems. For this reason, we believe that the current algorithm is more relevant for massive MIMO systems than that of the conventional MIMO systems.

## V. Simulation results

In this section, we provide simulation results. All of our results are obtained by averaging 20000 channel realizations and the WSMSEs are the normalized WSMSE (i.e., achieved WSMSE of (8) divided by $KM$). For this simulation, we consider a downlink multiuser massive MIMO system with $M = 128$, $K = 32$, $P_k = 1\text{mw}, \forall k$, the path losses are normalized to 1 (i.e., $0 < g_k < 1, \forall k$) and the signal to noise ratio (SNR) is defined as $\frac{P_{av}}{\sigma^2}$, where $P_{av} = \frac{1}{K} \sum_{k=1}^{K} P_k$. We also compare the performance of the proposed algorithm (i.e., Section IV) with that of the existing algorithm discussed in Section III.

Unless stated otherwise, we set the overall pilot symbol $\mathbf{X}$ as the first $K$ columns of the matrix $\bar{\mathbf{X}}$ for the existing algorithm, and we use this $\mathbf{X}$ as an initialized pilot symbols for the proposed algorithm, where $\bar{\mathbf{X}} = [\mathbf{U} \ \mathbf{U}]$ (i.e., with pilot reuse) and $\mathbf{U}$ is the normalized discrete Fourier transform (DFT) matrix of size $N$.

$$
\mathbf{G} = \begin{bmatrix}
0.0450 & 0.7400 & 0.8191 & 0.2608 \\
0.7040 & 0.2965 & 0.0823 & 0.8754 \\
0.0775 & 0.7410 & 0.1251 & 0.4437 \\
0.5925 & 0.6363 & 0.5327 & 0.2087 \\
0.6737 & 0.2419 & 0.7205 & 0.4000 \\
0.3940 & 0.4115 & 0.1497 & 0.8782 \\
0.0218 & 0.9238 & 0.6326 & 0.0669 \\
0.6327 & 0.7537 & 0.7697 & 0.0697
\end{bmatrix}. \quad (17)
$$

### A. *Effect of SNR*

In this simulation, we examine the effect of SNR on the performance of the proposed and existing algorithms. To this end, we set $N = 16$ and $g_k, \forall k$ are taken from a uniform random variable in between 0 and 1 which are given in (17) (i.e., vectorization of $\mathbf{G}$). With these settings, we plot the achieved normalized WSMSE as a function of SNR. As can be seen from Fig. 1, the performance of the proposed algorithm outperforms the existing algorithm for all SNR regions. Furthermore, for both algorithms, the normalized WSMSE decreases as the SNR increases which is expected.

### B. *Effect of the number of pilot symbols $N$*

In this simulation, we examine the effect of $N$ on the performance of the proposed and existing algorithms. Fig. 2 shows the normalized WSMSE of the proposed and the existing algorithm for different settings of $N$ and SNR when $g_k, \forall k$ are as in (17) (i.e., vectorization of $\mathbf{G}$). This figure also shows that the proposed algorithm achieves less normalized WSMSE compared to that of the existing one for all SNR regions when $N < K$. And as expected, both algorithms achieve the same normalized WSMSE when $N = K$, and their WSMSE decreases as $N$ increases.

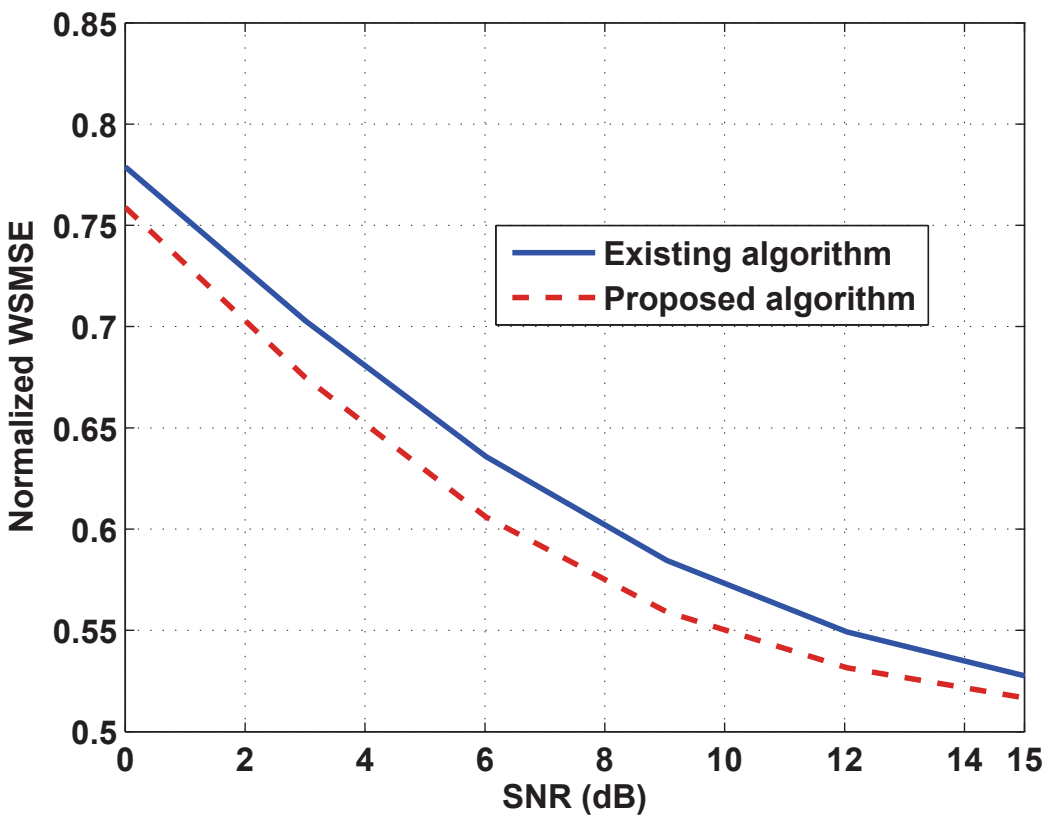

Fig. 1. Comparison of the proposed algorithm (**Algorithm I**) and the existing algorithm for $K = 32$ and $N = 16$

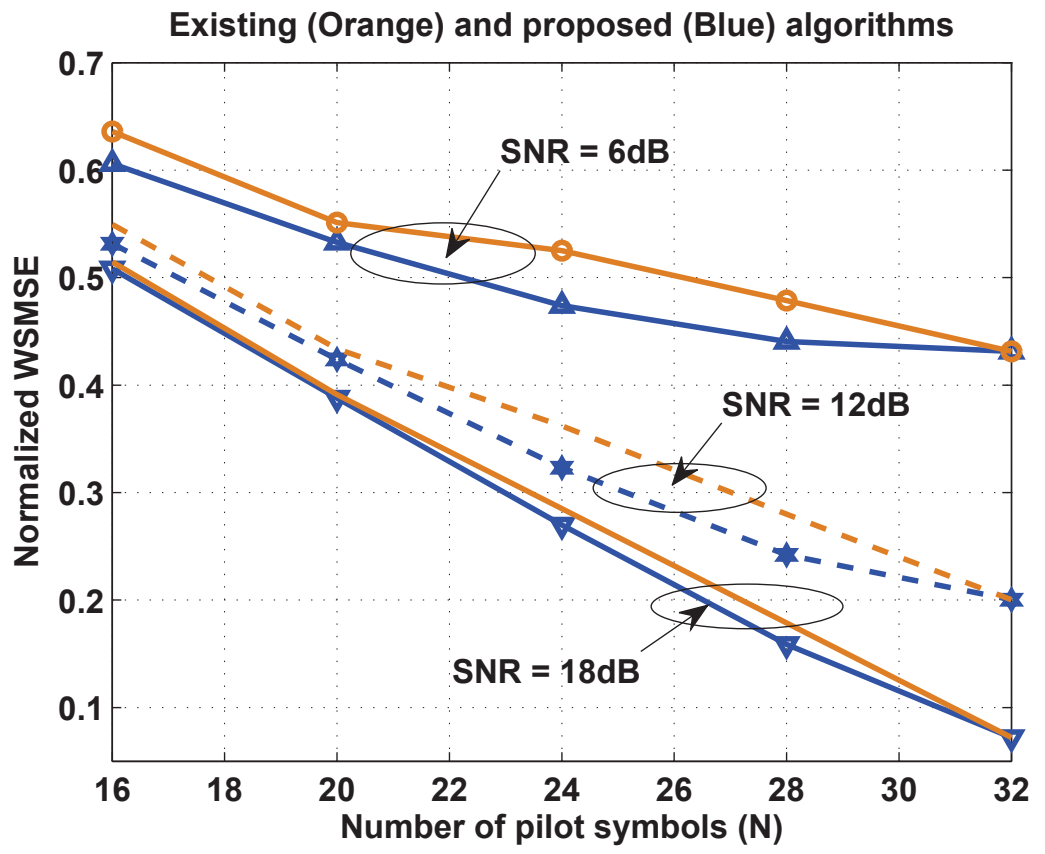

Fig. 2. Comparison of the proposed algorithm (**Algorithm I**) and the existing algorithm for $K = 32$ and different $N$.

### C. Convergence speed of the proposed algorithm

In this subsection, we study the convergence speed of the proposed algorithm. In this regard, we set $N = 16$ and use 3 initialization matrices. The first is like in Fig. 1, the second is the truncated DFT matrix of size $K$ and the third initialization is a random matrix taken from the complex Gaussian distribution with appropriate scaling. For these initializations, we plot the convergence behavior of the proposed algorithm in Fig. 3. As can be seen from this figure, faster convergence speed is achieved when the initialization is performed as a truncated version of the DFT matrix of size $K$. Nevertheless, for all initialization matrices, we have achieved the same normalized WSMSE in few iterations (i.e., $n = \frac{40}{K} < 2$ outer iteration is required for all initializations).

We would like to mention here that we have observed similar convergence behavior for other settings of $g_k, \forall k$, $\sigma^2$, $N$ and $K$.

## VI. Conclusions

In this paper, we propose novel pilot optimization and channel estimation algorithm for a multiuser massive MIMO system with $K$ single antenna MSs, arbitrary $N$ pilot symbols and TDD channel estimation method. The proposed algorithm is explained as follows. First, we formulate the channel estimation problem as a WSMSE minimization problem containing pilot symbols and introduced variables. Second, for fixed pilot symbols, the introduced variables are optimized using MMSE and generalized Rayleigh quotient methods. Finally, for $N = 1$ and $N = K$ settings, the pilot symbols are optimized using SDP convex optimization approach, and for the other settings of $N$ and $K$, the pilot symbols are optimized by applying simple iterative algorithm. When $N = K$, it is shown that the latter iterative algorithm gives the optimal pilot symbols achieved by the SDP method. Simulation results confirm that the proposed algorithm achieves less WSMSE compared to that of the conventional semi-orthogonal pilot symbol and MMSE channel estimation algorithm which creates pilot contamination.

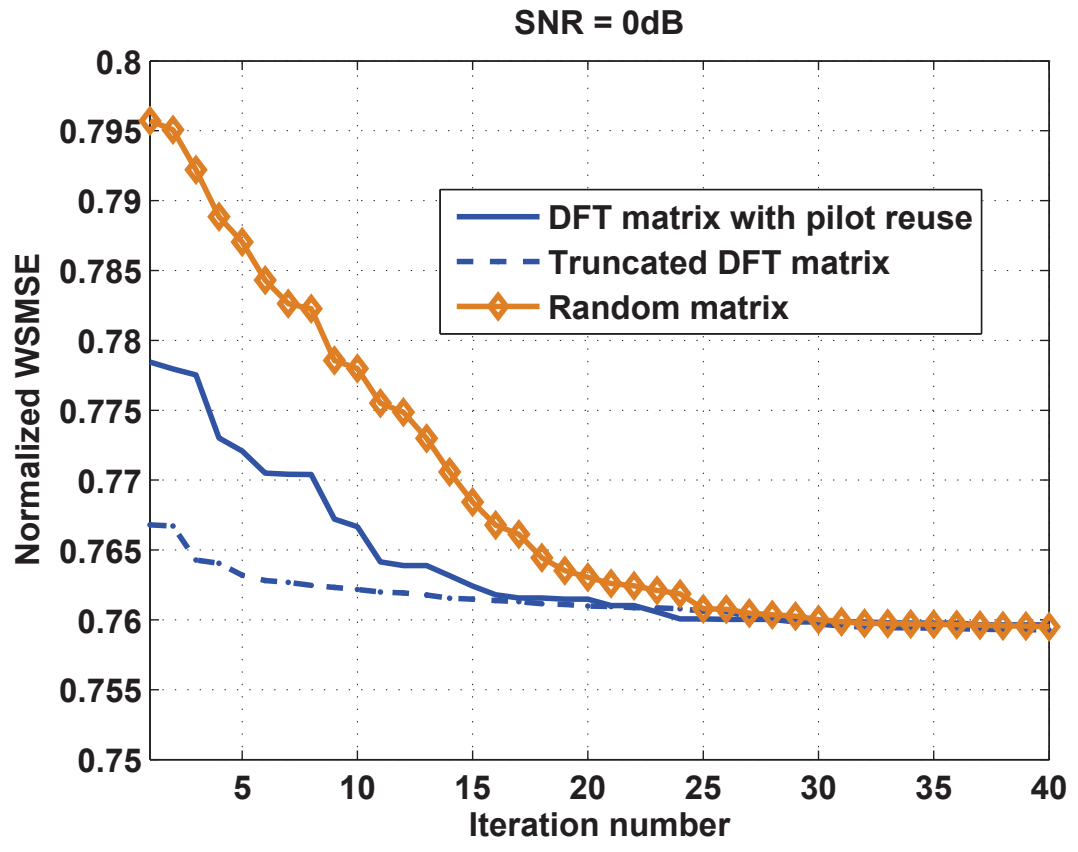

(a)

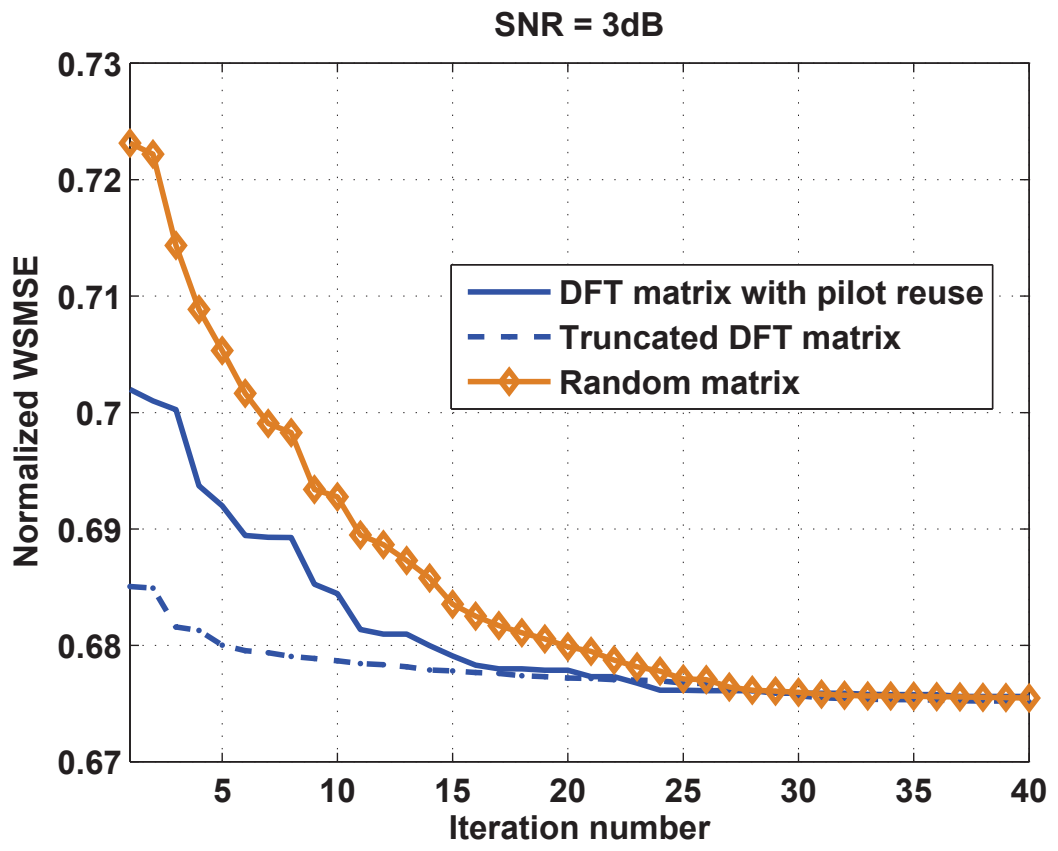

(b)

Fig. 3. Convergence speed of the proposed algorithm when $N = 16$ and $K = 32$ $[n = \frac{\text{Iteration no}}{K} \approx \frac{40}{32} < 2]$. (a) SNR = 0dB. (b) SNR = 3dB.